\documentclass[a4paper,twocolumn,fleqn]{article} % final: twocolumn 10pt
\usepackage{amsmath}                 % recommended for advanced math
\usepackage[psamsfonts]{amssymb}     % lots of math symbols (unnecessary with txfonts)
\usepackage{txfonts}                   % free Times-based fonts for text and math
\usepackage[pdftex]{graphicx}       % pdftex
\usepackage{tabularx}
\usepackage{pfr_APiP2025}                 % PFR style for APiP2025 paper

\begin{document}

\title{Benchmarking the Plane-Wave Born and Distorted Waves approximations for electron-impact collision strength computations: the sample case of Sr II}

%Electron-impact excitation collision strengths and forbidden transition radiative parameters in Sr II for nebular-phase kilonova modeling

%OU : Benchmarking the Plane-Wave Born and Distorted Waves approximations for electron-impact collision strength computations: the sample case of Sr II (MAIS alors on ne parle pas des forbidden transitions... ??)

% Use \sup{1},... for superscripts in the authors and affiliations lines.
\author{J\'er\^ome~Deprince\sup{1,2} and Lucas~Maison\sup{1}}

\affiliation{\normalsize %University Mons, Belgium \\
  \sup{1}Universit\'e de Mons (UMONS), Mons, Belgium \\
  \sup{2}Universit\'e Libre de Bruxelles (ULB), Brussels, Belgium}

%\date{(Received 1 December 2011 / Accepted 1 January 2012)}
\date{}

\email{Jerome.Deprince@umons.ac.be}

\begin{abstract}
The discovery of gravitational waves from a neutron star merger in 2017 (GW170817) and the associated kilonova (AT2017gfo) confirmed these events as key sites for heavy element production through the r-process. Subsequent observations, including late-time spectra with \textit{JWST}, have highlighted the need for accurate modeling of kilonova ejecta. In the photospheric phase, atomic level populations can be estimated under LTE using Boltzmann and Saha relations, but about a week after the merger the ejecta enters the nebular phase where non-LTE effects dominate. Modeling nebular spectra therefore requires a detailed treatment of radiative and collisional processes that affect the population of atomic levels.
This work focuses on electron-impact excitation in Sr II, a heavy ion relevant for kilonova spectra. Two computational approaches are employed: the Plane Wave Born approximation within the pseudo-relativistic Hartree-Fock method, and a Distorted Waves method using AUTOSTRUCTURE. The resulting collision strengths are compared against reference R-matrix data to evaluate the accuracy of these approximations and their suitability for large-scale applications to all heavy elements. In addition, radiative parameters for forbidden transitions are computed. These results provide an essential benchmark of approximations that could be used to compute atomic data for nebular-phase kilonova modeling.
\end{abstract}

\keywords{\normalsize kilonova, strontium, collision strengths, forbidden transitions, electron-impact excitation, nebular-phase, non-local thermodynamic equilibrium, neutron star merger, plane-wave Born, distorted waves}

%\DOI{10.1585/pfr.0.0000000}

%Cover page information
\AbstNum{I-20/A-11}
%\RegistrationNum{}
\CorrespondingAuthor{J\'er\^ome Deprince}
\CorrespondingAffiliation{University of Mons (UMONS)}
\PostalAddress{20 Place du Parc, 7000 Mons, Belgium}
\Tel{+32 65 37 36 28}
\Fax{/}

%\coverpage

%Cover page information
%\AbstNum{xxx (I-2, O-3, P-4, etc.)}
%\RegistrationNum{}
%\CorrespondingAuthor{Firstname~LASTNAME}
%\CorrespondingAffiliation{University of Plasma}
%\PostalAddress{322-6 Toki, Gifu 509-5292, Japan}
%\Tel{+81-572-58-0000}
%\Fax{+81-572-58-0001}

%\coverpage

\maketitle

\begin{normalsize}

\newpage
%%%%% MAIN TEXT %%%%%%%%%%%%%%%%%%%%%%%%%%%%%%%%%%%%%%%%%%%%%%%%%%%%%%%
\section{Introduction}
Gravitational waves originating from a neutron star merger (NSM) were observed for the first time in 2017 (GW170817) \cite{Abbott2017}. An electromagnetic signal known as a kilonova (KN) emitted by newly synthesized r-process elements during such an event was also detected (AT2017gfo), suggesting the presence of heavy elements in the KN ejecta \cite{Kasen2017}. In particular, lines from Sr II have been identified in the AT2017gfo spectrum \cite{Watson2019}, only heavy element that has undoubtedly been detected in a KN signal so far. More recently, the spectrum of a new KN was observed by the \textit{James Webb Space Telescope (JWST)} at later times, namely 29 and 61 days after the merger \cite{Levan2024}.

NSMs are promising candidate sites of heavy element production by rapid neutron capture \cite{Metzger2010}, a process that remains incompletely understood so far. KN modeling has thus become a topic of significant scientific interest, and substantial efforts have been made to model the atomic structure of heavy elements since then (e.g., \cite{Gaigalas2019,Tanaka2020, Fontes2020, Fontes2023, Flors2023, Deprince2025}).

During the first few days after the merger, the KN ejecta is in photospheric phase, in which the local thermodynamic equilibrium (LTE) assumption is valid \cite{Pognan2022}, simplifying the calculation of atomic energy level populations through Boltzmann and Saha equations. The photospheric-phase KN spectrum is greatly affected by a significant opacity due to the absorption of light by millions of electric dipole transitions (E1) from the various heavy ions present in the ejecta. Most of the above-mentioned works focused on heavy-elements LTE opacity calculations in this context. At later times, namely from about a week post-merger, the rapid decrease of the NSM ejecta temperature and density resulting from its expansion leads to non-local thermodynamic equilibrium (NLTE) conditions, corresponding to the KN ejecta nebular phase \cite{Pognan2022}, which makes the determination of energy level populations extremely complex. In addition, unlike the quasi-blackbody continuum spectrum observed in the photospheric phase, the nebular-phase spectrum is dominated by emission lines. In this phase, forbidden lines of magnetic (M1) and electric quadrupole (E2) types were presumably observed in the KN spectrum detected by the {\it Spitzer Space Telescope} \cite{Kasliwal2022}. Moreover, \cite{Gillanders2022} showed that the inclusion of heavy element M1 and E2 lines can lead to elemental signature identifications through radiative transfer simulations, while stressing the lack of available accurate atomic data for such lines in heavy elements.   

In this paper, we focus on atomic parameters in Sr II that are required for the KN nebular phase spectral analysis and modeling. This element was chosen since it was identified in the AT2017gfo KN spectrum and as atomic data already exist for comparison. In particular, we provide new radiative rates for two forbidden lines involving experimentally determined energy levels belonging to the ground configuration of Sr II. In order to do so, we used both the pseudo-relativistic Hartree-Fock method (HFR) as implemented in Cowan's code \cite{Cowan1981} and the multiconfiguration Breit-Pauli (MCBP) approach within the AUTOSTRUCTURE (AS) code \cite{Badnell2011}. In addition, we computed electron-impact excitation effective collision strengths for transitions involving the lowest energy levels of Sr II. To do so, the same above-mentioned methods were employed, in which the continuum electrons were modeled by means of two different approaches, namely the Plane-Wave Born (PWB) and the Distorted Waves (DW) approximations, respectively implemented in HFR and AS. They both consist (especially PWB) in poorer approximations compared to Close-Coupling (CC) R-Matrix computations \cite{Burke2011}, but calculations are much less computationally demanding than the latter. The main purpose of the present work is to benchmark such approaches with respect to more complex and accurate time-consuming calculations (as R-Matrix ones) in the sample case of Sr II \cite{Mulholland2024} to determine to what extent they can be used for large-scale computation purposes, in the same way as our large-scale opacity computations in all heavy elements \cite{Deprince2025}.

\section{Theoretical approaches}

\subsection{Pseudo-relativistic Hartree-Fock method}

In the HFR method, which is implemented in the Cowan's code \cite{Cowan1981}, a set of orbitals is obtained for each configuration included in the model by solving the coupled integro-differential Hartree-Fock (HF) equations, which arise from a variational principle applied to each configuration average energy. Based on the Slater-Condon theory, the atomic wavefunctions are built as superpositions of basis wavefunctions. Some relativistic corrections are also included. The HFR method is fully described in \cite{Cowan1981} and more details can be found in our previous papers (e.g., \cite{Deprince2025}). 

\subsection{Breit-Pauli multiconfiguration method (AUTOSTRUCTURE)}

In the MCBP method implemented in the AS code (described in details in \cite{Badnell2011}), the wavefunctions are built by using single-electron orbitals generated from a scaled Thomas-Fermi-Dirac-Amaldi potential. The scaling parameters for each orbital are optimized in a multiconfiguration variational procedure minimizing a weighted average of non-relativistic contributions to the energy. Spin-orbit coupling and Breit-Pauli operators are introduced as perturbations to obtain fine-structure relativistic
corrections.

\subsection{Electron-impact excitation process: the Plane-Wave Born and the Distorted Waves approximations}
A key parameters for electron-impact process modeling is the collision strength, $\Omega_{i\rightarrow j}$, which evaluates the likelihood of an energy transition from some initial level $i$ to some final level $j$ caused by the collision between a continuum electron and an ion. It is related to the cross section $\sigma_{i\rightarrow j}$ and defined as
\begin{equation}
    \Omega_{i\rightarrow j} = \frac{g_i\,k_i^2}{\pi\,a_0^2}\,\sigma_{i\rightarrow j},
\end{equation}
where $g_i$ is the statistical weight of the initial state, $k_i^2$ is the energy of the incident electron, and $a_0$ is the Bohr radius. Effective collision strengths ($\Upsilon_{i\rightarrow j}$) can also be obtained by Maxwellian averaging over a Boltzmann distribution of electron temperatures as
\begin{equation}
\Upsilon_{ij}(T_e) = \int_0^\infty  \Omega_{i\rightarrow j}\,e^{-\epsilon_j/kT_e}\,d\left(\frac{\epsilon_j}{kT_e}\right).
\end{equation}

In order to model collisional processes as the electron-impact excitation, the continuum electron can be described using various approximations, such as the PWB and the DW approaches. The PWB approximation neglects the interaction between the continuum electron wavefunction and the ionic target potential. This approximation is implemented in the HFR Cowan's code to model continuum electrons. An empirical correction arising from comparison to more complex DW and CC calculations is also computed \cite{Cowan1981} as \begin{equation}
    \Omega^{m2} = \Omega \left(X+\frac{3}{1+X}\right),
\end{equation}
where $X$ is the ratio between the continuum incident electron energy, $\epsilon$, and the transition energy, $\Delta E$, i.e. $X = \epsilon/\Delta E$.

In the DW approach, instead of plane waves (as in the PWB approximation), the wavefunctions of the incoming and outgoing electrons are modified by the scattering potential that accounts for the interaction between the continuum electron and the ionic target. In this approximation, the couplings between the various channels (thus, the resonances) are neglected, only the interaction between the initial and the final states is considered \cite{Badnell2011}.

\section{Atomic structure of Sr II}
In both methods (HFR and AS), the same multiconfiguration model was used. The latter is based on the configurations from which experimental levels are available at NIST \cite{NIST}, enriched by additional configurations with open 4p subshell, giving rise to the following list: 4p$^6$ $n$s with $5\leq n \leq 13$, 4p$^6n$p with $5\leq n \leq 8$, 4p$^6n$d with $4\leq n \leq 14$, 4p$^6n$f with $4\leq n \leq 12$, 4p$^6n$g with $5\leq n \leq 11$, 4p$^5$5s$^2$, 4p$^5$5s$n$s with $6\leq n \leq 8$, 4p$^5$5s$n$p with $5\leq n \leq 8$. A semi-empirical fit to the available experimental energy levels from \cite{Moore1952,Sansonetti2012} and available in the NIST database \cite{NIST} was also performed with HFR to obtain a better agreement and correct our radial parameter values. For AS, a Term Energy Correction (TEC) \cite{Badnell2011} was applied to optimize the energy levels.\\
Since we are interested in the KN ejecta nebular phase that is characterized by a relatively low temperature which does not exceed a few thousands Kelvins (or even much less), only the lowest energy levels can be populated. For example, \cite{Gillanders2025} assume that only levels whose energy is lower than 4 eV can be populated in the KN nebular phase, mentioning that this threshold was deliberately chosen quite high. For this reason, we limit our analyses to only the first five energy levels of Sr II in the present paper, since the sixth level lies above the threshold energy of 4 eV. A comparison between our energy levels computed by both HFR (without and with fit) and AS with respect to experimental values from NIST \cite{NIST}, as well as with the values computed by \cite{Mulholland2024} using AS is shown in Table \ref{tab:levels} for the first five levels of Sr II. Compared to the AS computation from \cite{Mulholland2024}, we obtained more accurate energy levels since the average agreement with the NIST values is of 0.02\% with HFR+Fit and of 0.10\% with AS, while results from \cite{Mulholland2024} show a difference of 5.29\% with NIST values on average. This can be explained by the different multiconfiguration models used but especially by the fit performed in the HFR calculations and by the TEC applied to the energy levels in the AS computation. As an example, the fit performed for the HFR values improves the average agreement from 0.87\% to 0.02\%.

\begin{table*}[]
    \centering
    \begin{tabular}{|c|c|c|c|c|c|c|c|}
    \hline
    Label & Configuration & Level & HFR (this work) & HFR+Fit (this work) & AS (this work) & AS \cite{Mulholland2024} & NIST \\
    \hline 
    1 & 4p$^6$5s & $^2$S$_{1/2}$ & 0     & 0     & 0     & 0     & 0     \\ 
    2 & 4p$^6$4d & $^2$D$_{3/2}$ & 14719 & 14558 & 14520 & 15686 & 14556 \\ 
    3 & 4p$^6$4d & $^2$D$_{5/2}$ & 15072 & 14832 & 14853 & 16001 & 14836 \\ 
    4 & 4p$^6$5p & $^2$P$_{1/2}$ & 23832 & 23706 & 23708 & 24378 & 23715 \\ 
    5 & 4p$^6$5p & $^2$P$_{3/2}$ & 24452 & 24514 & 24512 & 25192 & 24517 \\ 
    \hline
    \end{tabular}
     \caption{Comparison of our energy levels computed with HFR (without and with fit) and AS to experimental values from the NIST atomic database \cite{Moore1952,Sansonetti2012,NIST} and to AS calculations from \cite{Mulholland2024} in units of cm$^{-1}$.}
    \label{tab:levels}
\end{table*}

\section{Radiative parameters for forbidden transitions}
Two forbidden M1/E2 transitions are possible between the lowest metastable energy levels belonging to the ground configuration of Sr II, namely 4p$^6$4d $^2$D$_{3/2,5/2}$ -- 4p$^6$5s $^2$S$_{1/2}$. We computed their transition probabilities by means of both the HFR (with fit) and AS methods, using the above-mentioned multiconfiguration model. A comparison of our results with respect to the values obtained by \cite{Mulholland2024} and to experimental levels from \cite{Letchumanan2005,Biemont2000} and available in the NIST database \cite{NIST} is given in Table \ref{tab:forbidden}. We find that our AS transition rates underestimate the NIST values by about 10\%, while the results from \cite{Mulholland2024} overestimate them by 10\%. Nevertheless, our HFR transition probabilities are found to be much closer to the observed values from NIST, since the agreement is within 4\%, highlighting the accuracy of our HFR model and calculation (including a semi-empirical adjustment). 

\begin{table*}[]
    \centering
    \begin{tabular}{|c|c|c|c|c|c|}
    \hline
        Transition & Wavelength & HFR+Fit (this work)  & AS (this work) & AS \cite{Mulholland2024} & NIST   \\
        \hline
        4p$^6$4d $^2$D$_{5/2}$ -- 4p$^6$5s $^2$S$_{1/2}$ & 673.8392  & 1.58E+01 & 1.37E+01 & 1.75E+01 & 1.54E+01 \\
        4p$^6$4d $^2$D$_{3/2}$ -- 4p$^6$5s $^2$S$_{1/2}$ & 686.8171 & 9.59E+00 & 8.18E+00 & 1.05E+01 & 9.20E+00 \\
         \hline
    \end{tabular}
    \caption{Forbidden transitions involving the lowest metatable levels of Sr II computed by the HFR and AS methods, in comparison the those computed by \cite{Mulholland2024} using a CC approach and to the experimental values available in the NIST database REF. The observed wavelengths are given in nm, while all radiative rates are in units of s$^{-1}$.}
    \label{tab:forbidden}
\end{table*}

\section{Electron-impact excitation effective collision strengths}
We calculated the electron-impact excitation effective collision strengths for ten allowed and forbidden transitions between the five lowest energy levels of Sr II by means of both the HFR and the AS methods, respectively using the PWB and the DW approximations for the continuum electron treatment, for several temperatures below 10,000 K. The results obtained are shown in Table \ref{tab:collision} for the lowest temperature considered, i.e. $T=1000$ K, which is a typical temperature characterizing nebular-phase KN ejecta, and are compared to the effective collision strengths computed by \cite{Mulholland2024} using a more complex CC approach by means of the R-Matrix method. Figures \ref{fig:forbidden} and \ref{fig:allowed} show the evolution of the electron-impact excitation effective collision strengths with respect to the temperature for both our HFR-PWB and AS-DW results in the sample cases of two representative transitions, the first one being a forbidden transition while the second one is an allowed transition. In both case, they are compared to the R-Matrix calculations from \cite{Mulholland2024}, as well as to usual approximations used in most KN radiative transfer modeling code so far (in light of the lack of available atomic data, e.g. in \cite{Pognan2023}), namely the van Regemorter (for allowed transitions) \cite{VG1962} and the Axelrod formulae (for forbidden transitions) \cite{Axelrod1980}.\\
We can notice from Table \ref{tab:collision} that both HFR-PWB and AS-DW effective collision strengths only differ from the R-Matrix computations from \cite{Mulholland2024} by less than an order of magnitude. In most cases, they agree to each other within a factor of a few unities, with an only isolated exception for the effective collision strength computed by HFR-PWB for the forbidden transition 4p$^6$4d $^2$D$_{5/2}$ -- 4p$^6$4d $^2$D$_{3/2}$. The AS-DW results are in better agreement with the CC/R-Matrix computations from \cite{Mulholland2024} than the HFR-PWB approach (which was expected in light of the basic approximation the latter consists in), since they agree within a factor of $\sim$2.5 or less. Nevertheless, it is worth mentioning that the correct order of magnitude is obtained even with HFR-PWB (except for one isolated case). The latter can thus consists in a reasonable approximation in the context of large-scale computation in which completeness has to be favored over detailed accuracy, at least in a first step, in light of the current lack of a complete set of atomic data in all elements for such process and in such conditions/context. Indeed, HFR-PWB calculations are quite fast and easy to perform, especially since we already have the HFR targets for all elements from $Z=20$ to $Z=103$ from our recent work \cite{Deprince2025}. It is also worth highlighting that both our HFR-PWB and AS-DW collision strengths represent much better approximations than the Axelrod formula for forbidden transitions as shown in Figure \ref{fig:forbidden}, which is the approximation formula commonly used in nebular-phase KN radiative transfer modeling code (e.g., \cite{Pognan2023}). Indeed, this formula is known to generally significantly underestimate the collision strengths \cite{Pognan2025}, which can be confirmed from the results obtained in this work. In addition, the effective collision strength of the isolated transition for which the HFR-PWB fails to reproduce the R-Matrix results is still comparable (even slightly better) to the effective collision strength derived from the Axelrod formula. However, the van Regemorter formula seems to be a reasonable approximation for allowed transitions since the results are generally comparable to our HFR-PWB and AS-DW values, even if the former is in some cases of poorer quality as for the transition illustrated in Figure \ref{fig:allowed}. The HFR-PWB and AS-DW approaches are thus found to be a better alternative to these approximated formulae, especially in the case of the Axelrod approximation for forbidden transitions. 
\begin{table}[]
    \centering
    \begin{tabular}{|c|c|c|c|}
    \hline
    Transition & HFR-PWB & AS-DW & R-Matrix \cite{Mulholland2024} \\
    \hline
     1--2 & 3.24E+00 & 3.71E+00 & 3.03E+00 \\
     1--3 & 4.87E+00 & 5.47E+00 & 3.75E+00 \\
     1--4 & 2.24E+01 & 9.85E+00 & 4.57E+00 \\
     1--5 & 4.4E+00 & 1.82E+01 & 8.35E+00 \\
     2--3 & 9.50E-02 & 5.40E+00 & 1.36E+01 \\
     2--4 & 3.55E+01 & 1.37E+01 & 2.05E+01 \\
     2--5 & 6.86E+00  & 7.26E+00 & 6.61E+00 \\
     3--4 & 5.37E+00 & 3.71E+00 & 4.57E+00 \\
     3--5 & 6.24E+01 & 2.63E+01 & 3.66E+01 \\
     4--5 & 2.49E+00 & 1.24E+01 & 5.30E+00 \\
    \hline 
\end{tabular}
\caption{Electron-impact excitation effective collision strengths for the ten transitions between the first five lowest energy levels of Sr II at a temperature $T=1000$ K, computed by both the HFR-PWB and the AS-DW approaches, in comparison to the CC results obtained by \cite{Mulholland2024} using R-matrix.}
\label{tab:collision}
\end{table}

\begin{figure}[ht]
    \centering
    \includegraphics[width=1\linewidth]{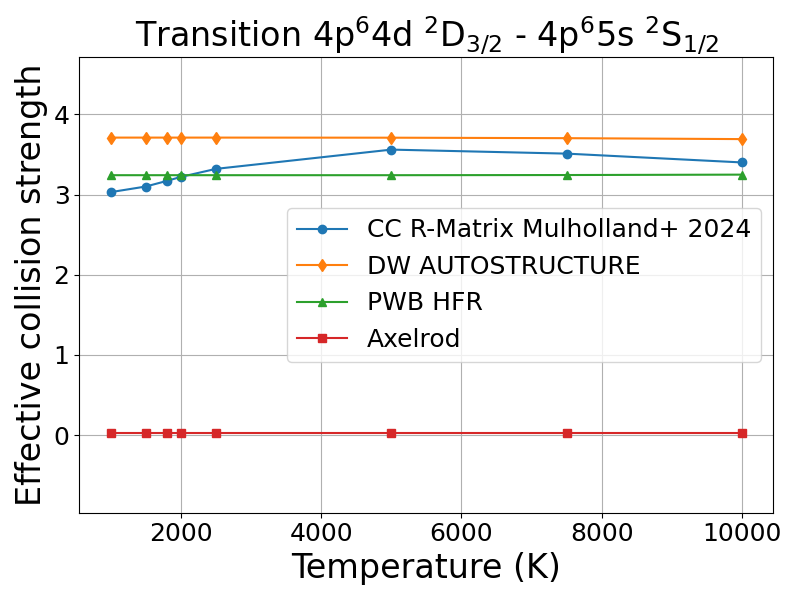}
    \caption{Effective collision strengths for a forbidden transition in Sr II computed with PWB-HFR and DW-AS as a function of the temperature, compared to R-Matrix results \cite{Mulholland2024} and to the Axelrod formula.}
    \label{fig:forbidden}

\end{figure}

\begin{figure}[ht]
    \centering
    \includegraphics[width=1\linewidth]{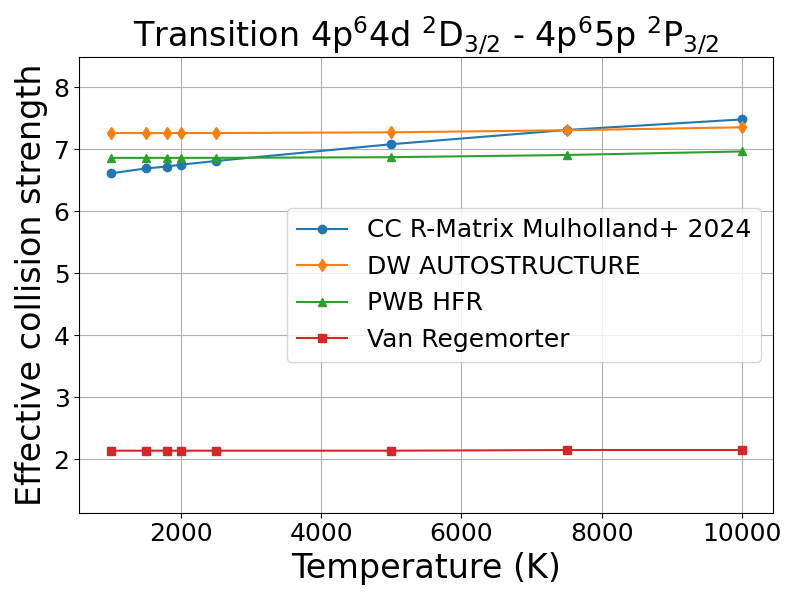}
    \caption{Effective collision strengths for an allowed transition in Sr II computed with PWB-HFR and DW-AS as a function of the temperature, compared to R-Matrix results \cite{Mulholland2024} and to the van Regemorter formula.}
    \label{fig:allowed}
\end{figure}

\section{Conclusions}
In this work, we have benchmarked two approximate methods—HFR-PWB and AS-DW—for computing electron-impact excitation effective collision strengths in Sr II against state-of-the-art CC/R-matrix data. While both approaches are intrinsically less accurate than close-coupling techniques, our results show that they generally reproduce the correct order of magnitude of effective collision strengths, with AS-DW yielding better agreement within a factor of 2.5. Importantly, even the simpler HFR-PWB method proves to be capable of capturing the relevant scale of excitation rates, which already represents a significant improvement over the empirical formulae commonly adopted in KN nebular-phase modeling codes. This is particularly true for forbidden transitions, where the Axelrod prescription can underestimate effective collision strengths by several orders of magnitude.\\
Given the acute lack of atomic data for the wide range of heavy elements expected in KN ejecta, the balance between completeness and accuracy becomes crucial. The present benchmark demonstrates that HFR-PWB and AS-DW approximations can provide reliable large-scale data sets of collisional parameters at relatively low computational cost, making them promising tools for systematic applications to lanthanides, actinides, and other heavy ions of astrophysical interest. The atomic data and methodology established here thus offer a practical pathway to improve non-LTE spectral modeling of nebular-phase KNe.

%\section*{Data availability statement}
%All the data from this work can be made available upon reasonable request to the corresponding author.  

\section*{Acknowledgments}
The present work is supported by the FWO and F.R.S.-FNRS under the Excellence of Science (EOS) programme (numbers O.0004.22 and O022818F)

\end{normalsize}

\end{document}